\documentclass[10pt,journal]{IEEEtranTCOM}

\usepackage[english]{babel}
\usepackage[usenames]{color}
\usepackage[cp1250]{inputenc}
\usepackage{amsfonts}
\usepackage{amsthm}
\usepackage{graphicx}
\usepackage{epsfig}
\usepackage{mathrsfs}
\usepackage{amsmath}
\usepackage{algorithm}
\usepackage{algorithmic}
\usepackage{hyperref}
\usepackage{tabularx}
\usepackage[centerlast,small,sc]{caption}

\theoremstyle{plain}

\begin{document}
\newcommand{\bea}{\begin{eqnarray}}
\newcommand{\eea}{\end{eqnarray}}
\newcommand{\be}{\begin{equation}}
\newcommand{\ee}{\end{equation}}
\newcommand{\beas}{\begin{eqnarray*}}
\newcommand{\eeas}{\end{eqnarray*}}
\newcommand{\bs}{\backslash}
\newcommand{\bc}{\begin{center}}
\newcommand{\ec}{\end{center}}
\def\SC {\mathscr{C}}

\title{Simple inexpensive vertex and edge invariants\\ distinguishing dataset strongly regular graphs}
\author{\IEEEauthorblockN{Jarek Duda}\\
\IEEEauthorblockA{Jagiellonian University,
Golebia 24, 31-007 Krakow, Poland,
Email: \emph{dudajar@gmail.com}}}
\maketitle

\begin{abstract}
While standard Weisfeiler-Leman vertex labels are not able to distinguish even vertices of regular graphs, there is proposed and tested family of inexpensive algebraic-combinatorial polynomial time vertex and edge invariants, distinguishing much more difficult SRGs (strongly regular graphs), also often their vertices. Among 43717 SRGs from dataset by Edward Spence, proposed vertex invariants alone were able to distinguish all but 4 pairs of graphs, which were easily distinguished by further application of proposed edge invariants. Specifically, proposed vertex invariants are traces or sorted diagonals of $(A|_{N_a})^p$ adjacency matrix $A$ restricted to $N_a$ neighborhood of vertex $a$, already for $p=3$ distinguishing all SRGs from 6 out of 13 sets in this dataset, 8 if adding $p=4$. Proposed edge invariants are analogously traces or diagonals of powers of $\bar{A}_{ab,cd}=A_{ab} A_{ac} A_{bd}$, nonzero for $(a,b)$ being edges. As SRGs are considered the most difficult cases for graph isomorphism problem, such algebraic-combinatorial invariants bring hope that this problem is polynomial time.
\end{abstract}
\section{Introduction}
In 2015 L{\'a}szl{\'o} Babai~\cite{babai} has shown that graph isomorphism problem can be solved in quasi-polynomial time ($2^{O((\log n)^c)}$ for some $c>0$), the question remains if it can be reduced to polynomial. The standard approach is Weisfeiler-Leman~\cite{WL} vertex labeling, not being able to distinguish even vertices of just regular graphs (of constant degree). To handle the most difficult cases like SRGs (strongly regular graphs)~\cite{cameron}, there is used transformation to graph with vertices as subsets of the original vertices, what corresponds to exponential cost growth. 

This article proposes and tests inexpensive polynomial time algebraic-combinatorial invariants, which directly (no transformation) distinguish all 43717 SRGs from dataset\footnote{used dataset: http://www.maths.gla.ac.uk/~es/srgraphs.php} by Edward Spence~\cite{spence1,spence2}. While it does not prove \textbf{completeness of this set of invariants}: that they can distinguish all non-isomorphic graphs (to show graph isomorphism is polynomial time), it brings hope for such proof in some future for some extended polynomial size set of discussed family of invariants.

\begin{figure}[t!]
    \centering
        \includegraphics[width=9cm]{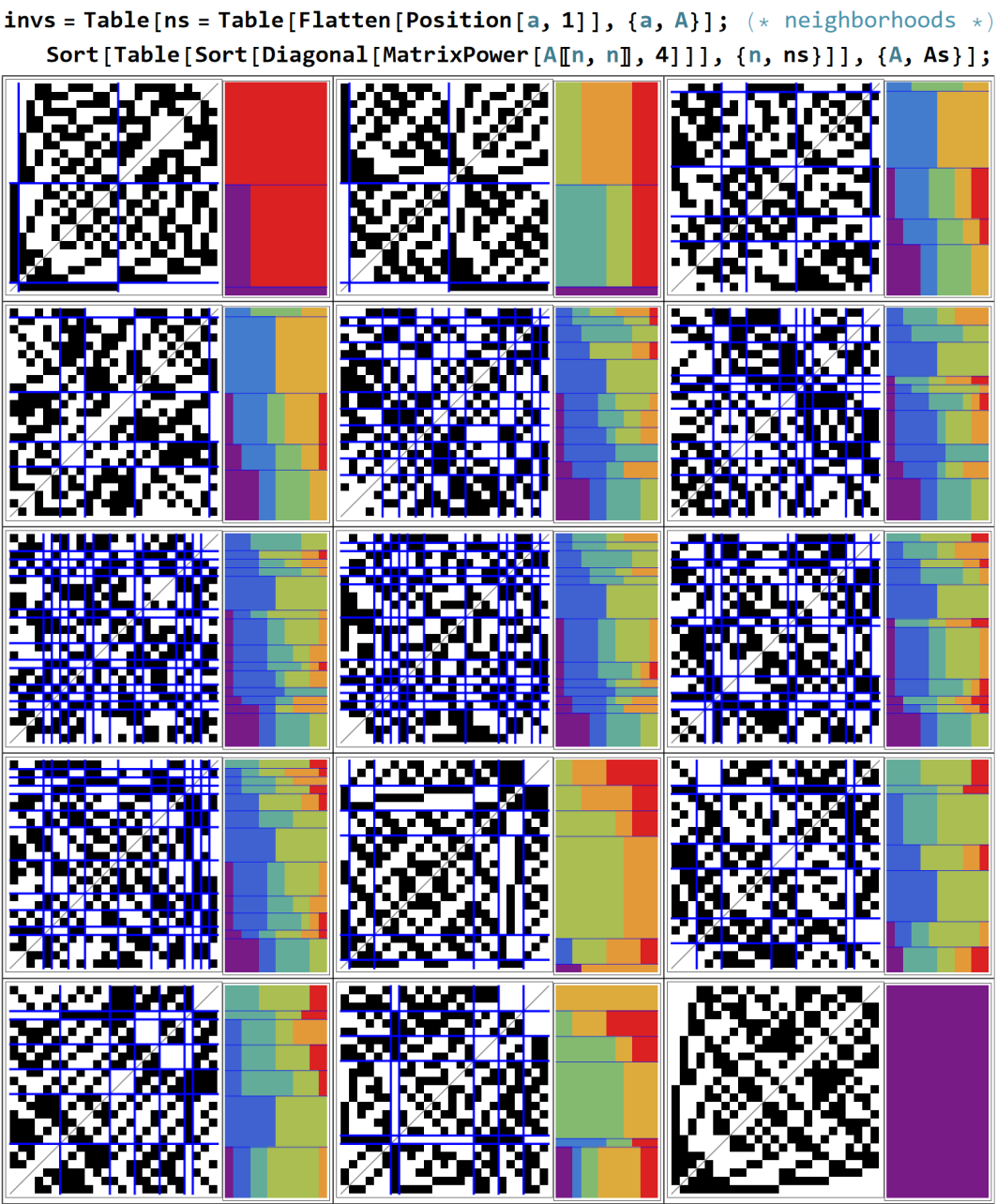}
        \caption{Example of proposed inexpensive \textbf{vertex invariants}: just sorted diagonals of $(A|_{N_a})^4$, combinatorially numbers of length 4 closed paths inside neighborhood of vertex $a$. There is shown used Mathematica code (top) and adjacency matrices on the left of sorted lexicographically visualized with colors calculated vertex invariants: $25\times 12$ matrices being graph invariants for 15 SRGs of 25-12-5-6 parameters, allowing to distinguish all of them as non-isomorphic. These vertex invariants usually allow to split vertices into subsets of constant invariants, marked by blue lines - allowing to restrict potential automorphisms to those applying permutations inside such subsets.
         }
        \label{d25}
\end{figure}

The basic versions of these invariants were proposed in algebraic form in \cite{duda2017p} focused on so far unsuccessful attempts for such proof. The current article improves them for practical applications (lower computational cost, extend trace to sorted diagonal), and actually test them on the largest dataset of SRGs the author were able to find - presenting the results in accessible way in multiple Figures and summarizing Table \ref{tab}, e.g. Fig. \ref{d25} showing vertex invariants distinguishing all 15 SRGs with 25-12-5-6 parameters.

\begin{table*}[t!]\centering 
\begin{tabular}{c|c||c|c|c|c|c|c|c||c|c|}
  \text{$v$-$k$-$\lambda$-$\mu$} & number & $\textrm{Tr}((A|_{N_a})^3)$ & $+\textrm{Tr}(()^4)$ & $+\textrm{Tr}(()^5)$ & $+\textrm{Tr}(()^6)$ & $+\textrm{Tr}(()^7)$ &   $+\textrm{Tr}(()^8)$  &   $+\textrm{Tr}(()^9)$ & same vert.&required\\
  parameters& of graphs &vertex invar.&vert. inv.&vert. inv.&vert. inv.&vert. inv.&vert. inv.&vert. inv.& invariants&edge inv.  \\
  \hline
 \text{16-6-2-2} & \textbf{2} & \textbf{2} & & &&&&&2&no\\ \hline
 \text{25-12-5-6} & \textbf{15} & 8 && &&&&&1&no \\ 
 sort(diag) &  & 13 & \textbf{15} &&&&&&& \\ \hline
 \text{26-10-3-4} & \textbf{10} & 8 & \textbf{10}&&&&&&0&no \\ \hline
 \text{28-12-6-4} & \textbf{4} & \textbf{4}&&&&&&&1&no \\ \hline
 \text{29-14-6-7} & \textbf{41} & 19 & 21 & &&&&&1&no\\
 sort(diag) &  &  \textbf{41}& &&&&&&&\\  \hline
 \text{35-18-9-9} & \textbf{3854} & 3359& 3722& 3741&3797&3798 && 3806&1&no  \\
 sort(diag) &  & 3798 & 3830 && 3847 &&&&& \\  \hline
 \text{36-14-4-6} & \textbf{180} & 86 & 161 &  165& 166 & 172 &177&&4&1 outblock\\ 
  sort(diag) &  &  88 & 176 & & 177&&&&&\\\hline
 \text{36-15-6-6} & \textbf{32548} & 21497 & 31645 & 31977 & 32314 & 32354& 32357 & 32378&4&1 outblock\\ 
 sort(diag) &  & 31321 & 32445 & 32497 &32510 &32511&&&& \\ \hline
 \text{37-18-8-9} & \textbf{6760} & 3300 & 3381 & &&&&&1&no \\ 
 sort(diag) &  &  \textbf{6760}&&&&&&&& \\\hline
 \text{40-12-2-4} & \textbf{28} & 23 &  &  & 26 &&&& 2&1 inblock \\\hline
 \text{45-12-3-3} & \textbf{78} & \textbf{78}&&& &&&&1&no\\ \hline
 \text{50-21-8-9} & \textbf{18} & \textbf{18}&&& &&&&0&no\\ \hline
 \text{64-18-2-6} & \textbf{167} & 134 &  & &151& &&&8&1 inblock \\ 
  \hline
\end{tabular}
\caption{Result summary for SRGs from Edward Spence dataset for 13 parameters (first column), all having more than one graph - their numbers are written in the second column. The following seven columns show numbers of distinguished subsets of graphs for the shown discussed vertex invariants using up to the written power - there are only shown changes of values, marked with bold if reaching the number of graphs for given parameter - distinguishing all of them. While the basic discussed approach is using traces of powers of $A|_{N_a}$ being sum of diagonal terms, we can alternatively sort these diagonal values, sometimes allowing to distinguish more vertices/graphs - in such cases there are shown two rows for given parameters: upper for use of trace, and lower "sort(diag)" for using sorted diagonal instead. Last two columns show the number of graphs with identical vertex invariants shown in Fig. \ref{unique}, and 4 pairs of graphs for which vertex invariants were insufficient to distinguish them shown in Fig. \ref{problem} - further edge invariants distinguished them. To reach only 4 such problematic cases, vertex invariants were also compared for subgraph with removed first block of distinguished vertices. }
   \label{tab}
\end{table*}

\section{Basic definitions and strongly regular graphs}
This Section briefly introduces basic tools, to prepare for proposed methods in the central next Section.

We work on graphs on $v$ vertices, given by $A\in \{0,1\}^{v\times v}$ adjacency matrix defining edges as  $E=\{(a,b):A_{ab}=1\}$. For simplicity assume these graphs are undirected ($A=A^T$) and there are no self-loops ($\forall_a A_{aa}=0$). Define neighborhood of vertex $a$ as $N_a=\{b:A_{ab}=1\}$.

Graph is called \textbf{regular} if all vertices have the same degree: $\forall_i \sum_j A_{ij} = k$. The standard Weisfeiler-Leman vertex labeling tries to distinguish vertices based on their degrees - unsuccessful already for regular graphs.\\

Much more difficult to distinguish are \textbf{strongly regular graphs} (SRGs)~\cite{cameron}: with $v$-$k$-$\lambda$-$\mu$ parameters - regular (constant degree $k$) and additionally:
\begin{itemize}
  \item every two adjacent vertices have $\lambda$ common neighbours,
  \item every two non-adjacent vertices have $\mu$ common neighbours.
\end{itemize}
Their adjacency matrix $A$ has to satisfy:
\be A^2 =kI +\lambda A + \mu(J+I-A)\qquad\textrm{for}\quad\forall_{ab}\, J_{ab}=1 \ee
leading to always only 3 different eigenvalues of $A$: one trivial 1D eigenspace $A(1,1,\ldots,1)=k (1,1,\ldots,1)$, and two large degenerate eigenspaces for eigenvalues:
\be \frac{1}{2}\left((\lambda-\mu)\pm \sqrt{(\lambda-\mu)^2+4(k-\mu)}\right)  \ee
Search for SRGs is far nontrivial, there is used the largest found dataset, summarized with results in Table \ref{tab}.

\section{Proposed graph isomorphism invariants}
This Section starts with algebraic motivation of the discussed invariants as in \cite{duda2017p}, then focus on tested practical vertex and edge invariants.

\subsection{The road to the discussed invariants} 
Let us start with algebraic motivation, introduction. The basic observation is well known \textbf{matrix similarity test}:
\be \forall_{p=1}^v \textrm{Tr}(A^p)=\textrm{Tr}(B^p)\ \Rightarrow\ \exists_{O:OO^T=I}\ A=OBO^T \ee
In polynomial time it allows to test existence of orthogonal matrix $O$ transforming between two matrices. Testing existence of graph isomorphism seems very similar, with restriction to $O$ being a permuation matrix - what can be characterized as orthogonal with 0/1 coefficients. This way the remaining question is: \textbf{among similarity matrices between $A$ and $B$, is there a permutation?}
\be \mathbf{O}_{AB}:=\{O:OO^T=I, A=OBO^T\}\quad \textrm{contain permutation?}\ee

SRGs show the real difficulty of this question: due to high degeneracy of the two nontrivial eigenspaces, such set of possible similarity matrices contains all basis rotations inside these two eigenspaces. It allows to formulate the question of existence of isomorphisms between two SRGs as a question if there exists a rotations between two sets of points, constructed from transposed bases of corresponding eigenspaces.\\

Analogously to $\textrm{Tr}(A^p)=\textrm{Tr}(B^p)$ tests, we would like to propose further polynomial time invariants conserved by permutations, hopefully to restrict $\mathbf{O}_{AB}$ to permutations only. Ideally we would like to find a polynomial size complete set of invariants - such that agreement on it would ensure existence of permutation/isomorphism.

While search for proof of such restriction to permuations alone was not successful so far, promising and experimentally successful direction from \cite{duda2017p} is going to tensor products, e.g. 
\be \hat{A}_{abc}:=\sum_d A_{ad} A_{bd} A_{cd} \ee
especially thanks to \textbf{uniqueness of tensor decomposition theorem}~\cite{tens} - allowing to conclude existence of sought permutation/isomorphism if only there exists orthogonal matrix $O$ between such rank-3 tensors: 
\be \exists_{O:OO^T=I}: \hat{A}_{abc}=\sum_{ijk}\hat{B}_{ijk}\,O_{ia}\,O_{jb}\,O_{kc}\quad ? \ee

This way we need to extend the $\textrm{Tr}(A^p)=\textrm{Tr}(B^p)$ similarity test from rank-2 (matrices) to rank-3 tensors. Unfortunately it is surprisingly difficult, more than e.g. analogously considering characteristic polynomial using tensor hyperdeterminant~\cite{hyper}. The main reason is dimensionality growth: intuitively, removing rotation information ($v(v-1)/2$ dimensions) from symmetric matrix ($v(v+1)/2$ dimensions) there remains rank-1 set of eigenvalues ($v$ dimensions). However, removing it from rank-3 tensor ($O(v^3)$ dimensions), there still remain $O(v^3)$ parameters.

Fortunately there is a general way to construct such rotation invariants: building some larger matrices (e.g. $v^2\times v^2$ e.g. discussed further (\ref{inveq})) from copies of $A$ and summation over intermediate vertices like in matrix power, and for such larger matrices test traces of powers similarity conditions.

While adjacency matrix modulo permuation should be completely determined by $O(v^2)$ independent invariants, in polynomial time we could construct e.g. $v^{10}$ invariants - naively much more than required. However, most of them would be rather dependent (like $\textrm{Tr}(A^{v+1})$), the big question is \textbf{how to prove that there is a sufficient number of independent among them?} In other words: that they form a complete set of invariants - fully determining matrix modulo permutation, distinguishing any non-isomorphic graphs.

\begin{figure}[t!]
    \centering
        \includegraphics[width=9cm]{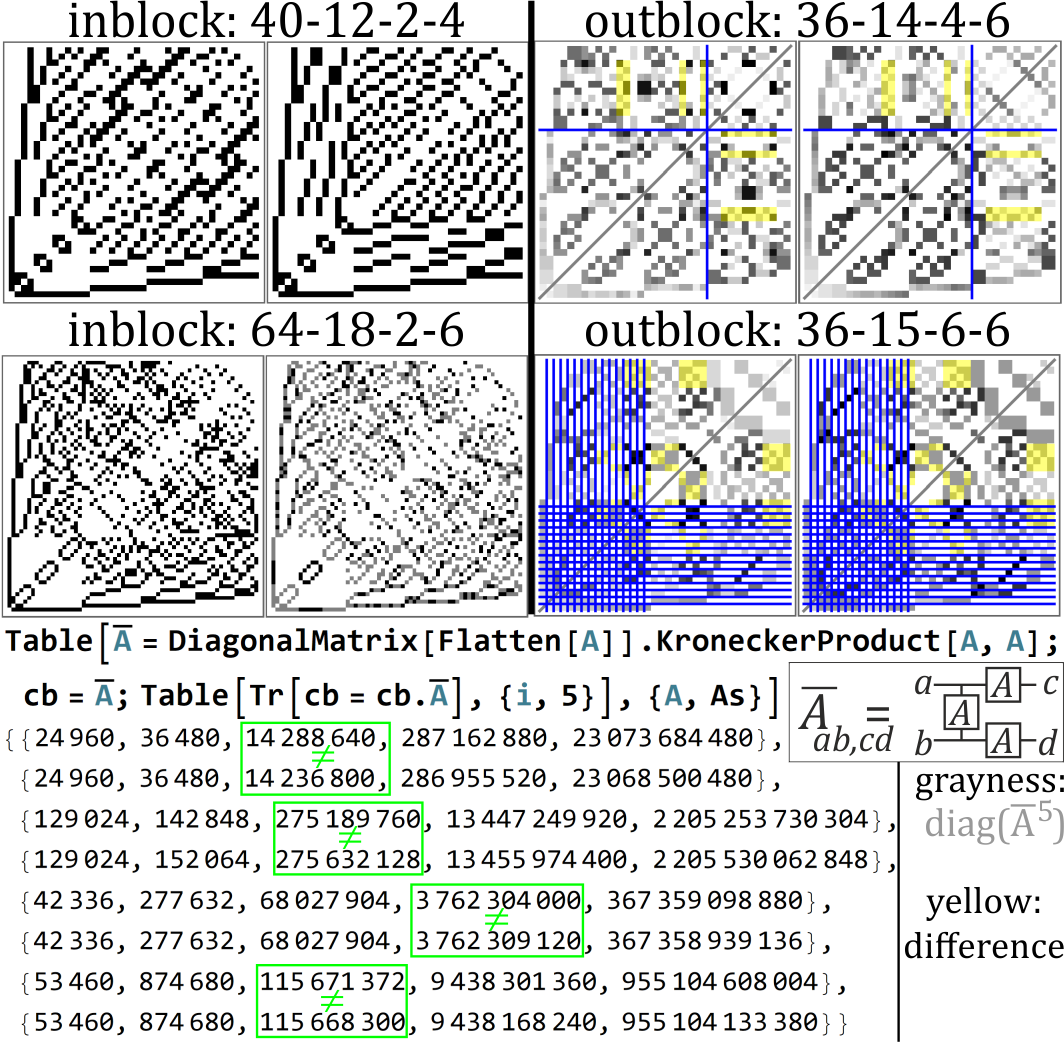}
        \caption{The only 4 pairs of graphs in the dataset for which the discussed vertex invariants were insufficient to distinguish - in contrast to the presented further edge invariant. For "inblock" two of them (left) the vertex invariants were not able to split vertices into subsets (happened to 39 of 43717 SRGs shown in Fig. \ref{unique}), and such invariants were identical for both graphs. For "outblock" two of them (right), beside sorted $\textrm{diag}((A|_{N_a})^p)$ vertex invariants for the entire graph, if needed there were also used such invariants for subgraph with removed first set of distinguished vertices - leaving the shown two pairs, having subtle differences of connections between its subgraphs (marked yellow). Grayness shows split of edges into subsets of constant values of $\textrm{diag}(\bar{A}^5)$.}
        \label{problem}
\end{figure}

\subsection{Practical vertex and edge invariants}
While there is no proof of completeness so far, here we focus on the simplest invariants from \cite{duda2017p} - improve them for practical applications, and test on the SRG dataset. Specifically, originally we build larger $v^2 \times v^2$ matrices:
\be \tilde{A}_{ab,cd} =A_{ab} A_{ac}\delta_{bd}\qquad\qquad \bar{A}_{ab,cd} =A_{ab} A_{ac}A_{bd} \label{inveq}\ee
and analogously for $\tilde{B}$, $\bar{B}$, then test similarity: if
\be \forall_{p=1,\ldots, v^2}\quad \textrm{Tr}(\tilde{A}^p)=\textrm{Tr}(\tilde{B}^p),\quad \textrm{Tr}(\bar{A}^p)=\textrm{Tr}(\bar{B}^p)\quad?\ee
For the tested SRG dataset, positive answer to this question allows to conclude being non-isomorphic in polynomial time.\\

Let us now discuss and optimize them for practical applications:
\subsubsection{Vertex invariants from $\tilde{A}$}
Trace of a matrix is sum of its diagonals, but observe we can sort this diagonal instead of summation - still being invariant under permutation, getting a larger number of invariants from the same matrix. The $\delta_{bd}$ in (\ref{inveq}) definition of $\tilde{A}$ makes that powers of this matrix can have nonzero coefficients still only when  $b=d$. The $A_{ab}$ in this definition enforces that $a$ in nonzero coefficients has to be in $N_b$ neighborhood of $b=d$.

This way a diagonal term: $(\tilde{A}^p)_{ab,ab}$ combinatorially contains the number of closed length $p$ paths starting and ending in $a$, all inside $N_b$ neighborhood of $b$.

This observation allows to work on powers of much smaller (size degree of $b$) adjacency matrix restricted to neighborhoods: $A|_{N_b}$, separately for each vertex $b$ to get its invariants. 

We can use both $\textrm{Tr}((A|_{N_b})^p)$, or more detailed sorted $\textrm{diag}((A|_{N_b})^p)$ - as we can see in Table \ref{tab}, which sometimes can provides essentially better distinction. 

Such vertex invariants for multiple powers $p$ can be concatenated into a longer vector, however, in practice the highest power usually contained information also about the lower ones - it seems sufficient to just use the highest calculated power (using e.g. $M^8=((M^2)^2)^2$). While to ensure similarity we should test for all the powers up matrix size (degree $k$ here), in this Table we can see that much lower powers like 3 or 4 are usually sufficient. 

Such vertex invariants usually allow to split the vertices into blocks of identical invariants - Fig. \ref{unique} shows all 39 (from 43717) SRGs for which such split was not successful. Automorphisms and isomorphism have to conserve such vertex invariants - can only permutate inside such blocks.

\begin{figure*}[t!]
    \centering
        \includegraphics[width=18cm]{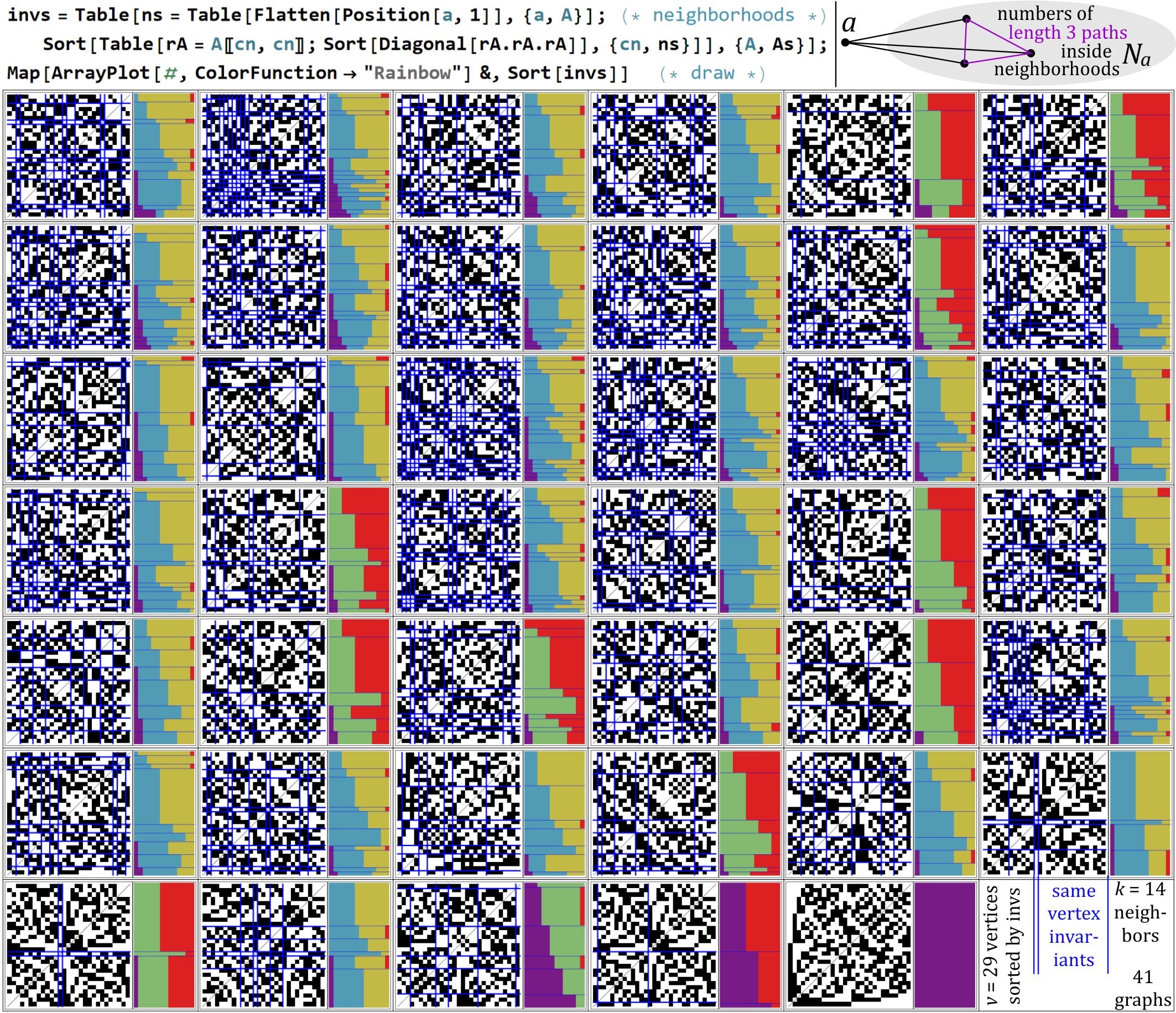}
        \caption{Analogously as in Fig. \ref{d25}, but for $p=3$ power and 41 SRGs of 29-14-6-7 parameters - allowing to distinguish them with sorted  $\textrm{diag}((A|_{N_a})^3)$ vertex invariants, also distinguish some vertices: splitting them into blocks with shown blue lines.}
        \label{diag}
\end{figure*}

To get graph invariants we can just sort lexicographically such vertex invariants - Figures \ref{d25} and \ref{diag} show two examples where for low single power it allowed to distinguish all SRGs for given parameters, also their found blocking into subsets of the same vertex invariants. 

However, in Table \ref{tab} we can see that in many cases such invariants were far from sufficient to distinguish all. As in most cases we have some blocking of vertices based on invariants, we can further compare these blocks - just removing vertices from the first block and testing the same invariants for such smaller adjacency matrix was sufficient to distinguish all but the 4 graph pairs in Fig. \ref{problem} - easily distinguished by below edge invariant.\\

\subsubsection{Edge invariants from $\bar{A}$}
While to ensure similarity we would have to test all powers up to the number of edges, in Fig. \ref{problem} power $p=4$ or 5 has turned out sufficient to distinguish the most problematic graphs.

Looking at diagonal of powers of $\bar{A}_{ab,cd} =A_{ab} A_{ac}A_{bd}$, due to $A_{ab}$ term they can be nonzero only for $(a,b)$ being an edge - to reduce computational cost, in practice we can restrict this matrix to $(a,b)$ and $(c,d)$ being edges of the considered graph.

Diagonal term $(\bar{A}^p)_{(a,b),(a,b)}$ for edge $(a,b)$ is combinatorially the number of length $p$ paths of edges, where two edges are neighboring when their ends are neighboring. 

\begin{figure*}[t!]
    \centering
        \includegraphics[width=18cm]{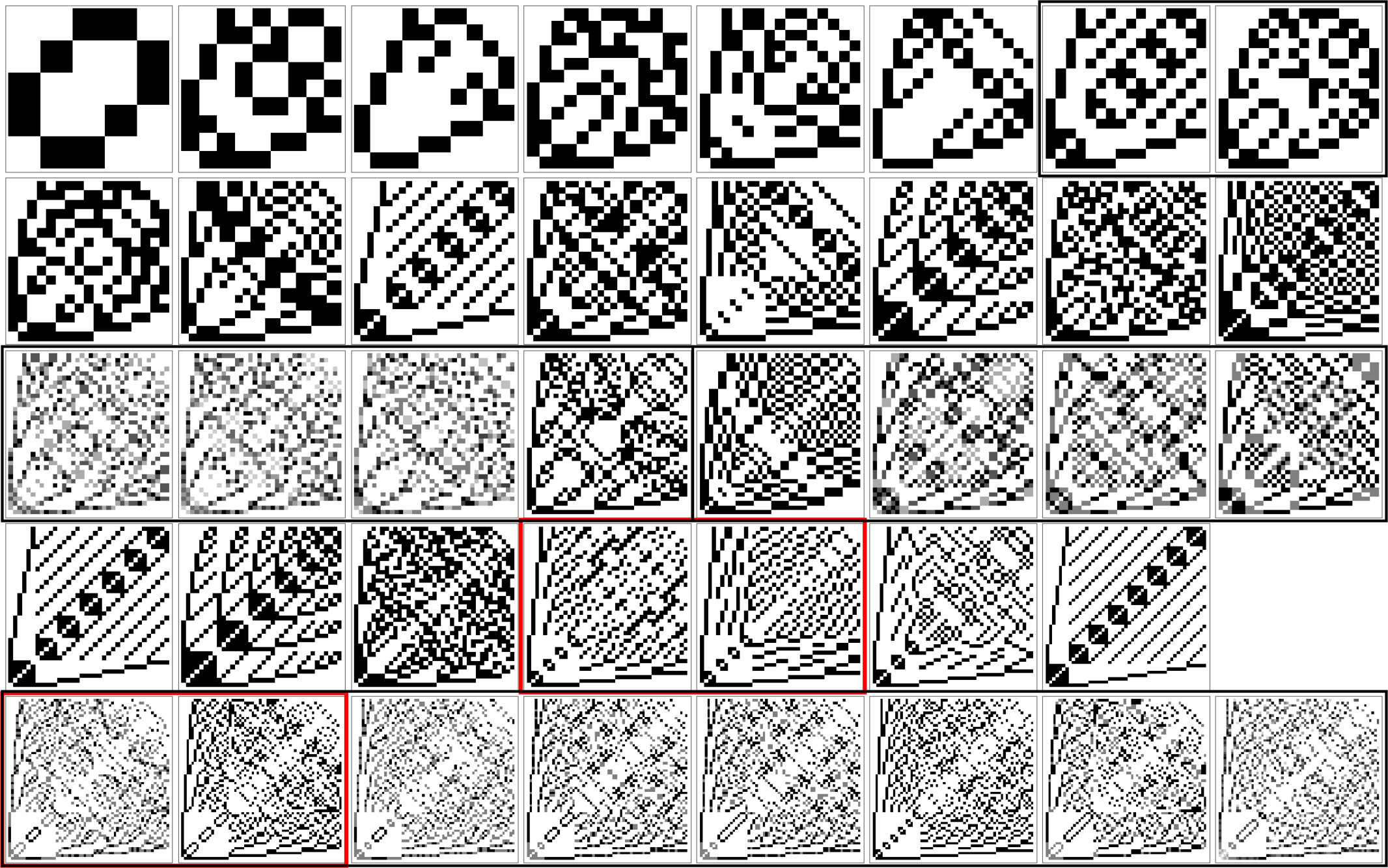}
        \caption{Adjacency matrices of all 39 SRGs in the dataset (of 43717 SRGs) for which the discussed vertex invariants were not able to distinguish vertices (e.g. to restrict automorphisms, they can still  differ between graphs allowing to distinguish graphs). Marked two pairs, shown also in Fig. \ref{problem}, are the only two requiring further edge invariants to be distinguish - they could also split edges into conserved subsets of the same invariants, marked with grayness.}
        \label{unique}
\end{figure*}

In trace $\textrm{Tr}(\bar{A}^p)$ we sum these diagonals - sufficient to distinguish the 4 problematic cases in Fig. \ref{problem}. Additionally, we can group edges having identical values in this diagonal of $\bar{A}^p$, which have to be conserved by isomorphism, and mark them as different types of edges - sometime being able to split edges into subsets conserved by isomorphism, marked with grayness levels in Figures \ref{problem}, \ref{unique}.

\section{Conclusions and further work}
There were proposed and tested simple inexpensive algebraic-combinatorial invariants, easily distinguishing (43717) available SRGs, being the most difficult cases of the graph isomorphism problem - suggesting both search for its practical application, and for formal proof that graph isomorphism is polynomial time.\\

Some possible future work direction directions:
\begin{itemize}
\item Development toward formal proof that graph isomorphism is polynomial - choose a set of invariants e.g. including the discussed ones and a polynomial number of others, such that completeness can be proven: that together they can distinguish any non-isomorphic graphs.
  \item Development toward practical applications - comparison with used state-of-art used methods, also for general matrices (not adjacency), maybe developing some heuristics (e.g. neural networks) guessing which invariants/powers are sufficient for various applications, for large powers use multiplication modulo to avoid large arithmetics, etc.
  \item Consider different promising invariants e.g. vertex from diagonal of powers $t(A)_{ab}=\sum_{cd} A_{ac} A_{ad} A_{cd} A_{cb} A_{bd}$, or $\tilde{A}^q_{ab,cd} =(A^q)_{ab} A_{ac}\delta_{bd}$ type extensions to handle graphs with chains of edges - which will be required to distinguish all non-isomorphic graphs.
  \item Use such invariants to better understand families of graphs like SRGs, maybe try to generate graphs additionally satisfying some of them, e.g. with vertices or edges indistinguishable this way like in Fig. \ref{unique}, or non-isomorphic SRGs made of identical subgraphs like "outblock" in Fig. \ref{problem}.
\end{itemize}

\bibliographystyle{IEEEtran}
\bibliography{cites}

\end{document}